\documentclass[doublecol]{epl2} 
\usepackage{amsmath}
\usepackage{graphics}

\newcommand{\vectr}[1]{\mathbf{#1}}

\title{Tunable transient decay times in nonlinear systems: application to magnetic precession}

\author{M. G. Phelps \and K. L. Livesey \and A. M. Ferona \and R. E. Camley}

\institute{
Department of Physics and Energy Science, University of Colorado at Colorado Springs - Colorado Springs, CO 80918
}

\abstract{
The dynamical motion of the magnetization plays a key role in the properties of magnetic materials. If the magnetization is initially away from the equilibrium direction in a magnetic nanoparticle, it will precess at a natural frequency and, with some damping present, will decay to the equilibrium position in a short lifetime. Here we investigate a simple but important situation where a magnetic nanoparticle is driven non-resonantly by an oscillating magnetic field, not at the natural frequency. We find a surprising result that the lifetime of the transient motion is strongly tunable, by factors of over 10,000, by varying the amplitude of the driving field.
}

\begin{document}

\maketitle

\section{Introduction}
A wide variety of fascinating topics have been studied in nonlinear magnetism \cite{b.1,b.2,b.3} and other branches of nonlinear physics. These include solitons \cite{b.4,b.5,b.6,b.7}, period doubling \cite{b.8}, nonlinear combinations of frequencies \cite{b.9,b.10}, strange attractors, limit cycles, chaos \cite{b.11}, and routes to chaos through bifurcation processes \cite{b.12,b.13}. The field has recently been invigorated by the use of nanostructures, which have allowed very large microwave fields to be applied, with amplitudes in excess of several hundred Oe \cite{b.14,b.15}, and have allowed interesting nonlinear conversion between modes \cite{b.16,b.17,b.18}.  Almost all of these topics deal with what is happening in the system after all transients have disappeared \cite{b.19,b.20}.
\\ \indent In this paper, in contrast, we theoretically examine the \textit{transient} precessional behavior appropriate for a magnetic nanoparticle. We find a surprising result that, in the nonlinear limit, the transient lifetime can be significantly extended by adding a strong oscillating driving field which is at a \textit{different} frequency than the natural frequency. This increase of the lifetime depends sensitively (and nonmontonically) on the strength of the driving field.  Near a critical driving field the lifetime can be extended by factors of over 10,000. We note that this behavior occurs in a nonlinear regime, but chaos is not required for this stabilization of the transient.
\\
\section{Transient magnetization dynamics: analysis and characterization}\subsection{Equations of motion}With a static magnetic field applied to the nanoparticle, there is a natural precessional frequency for the magnetization.  If the magnetic moment has an initial position, away from equilibrium, the system will exhibit transient behavior in that the magnetization will precess and at the same time decay toward its equilibrium direction as a result of magnetic damping.  The theoretical calculations start with the equations of motion for a macrospin, appropriate, for example, for a small magnetic particle with a diameter smaller than the exchange length so that the entire system is in a single domain. For instance, the exchange length in iron is about 2 nm.  The equation of motion for the magnetization~\cite{b.21}, $\vectr M$, is given by
\begin{equation}
\label{e.1}
\frac{\upd \vectr M}{\upd t}=\frac{\gamma}{1+\alpha^2}\left(\vectr M\times\vectr H - \alpha\frac{\vectr M \times (\vectr M \times \vectr H)}{M}\right)~.
\end{equation}
Here $\gamma$ is the gyromagnetic ratio, $\alpha$ is a dimensionless damping parameter, and the magnetic field, $\vectr H$, is a sum of a static field in the $\hat{\vectr z}$ direction, demagnetizing fields, and a driving field in the $\hat{\vectr x}$ direction, oscillating with a frequency $\omega$, namely
\begin{equation}
\label{e.2}
\vectr H = H_0~\hat{\vectr z} - \vectr N \cdot \vectr M + h_d \cos{\omega t}~\hat{\vectr x}~.
\end{equation}
We use a demagnetizing tensor appropriate to a sphere or a cube given by
\begin{equation}
\label{e.3}
\vectr N_{\alpha\beta} = \frac{4\pi}{3} \boldsymbol{\delta}_{\alpha\beta}~
\end{equation}
where $\boldsymbol{\delta}_{\alpha\beta}$ is the second rank Kronecker delta tensor.
\\
\subsection{Limiting cases}
In the absence of damping or a driving field the natural frequency of precessional motion about the $z$-axis is
\begin{equation}
\label{e.4}
\omega_0 = \gamma H_0~.
\end{equation}
As we will see, the transient precession will occur at this frequency in the linear limit, but will be shifted down slightly in the nonlinear case. Applying a driving field at this frequency can speed-up magnetization reversal in a nanoparticle \cite{b.22}.

In the absence of a static field and damping, the motion of the magnetization has an analytical solution.  We start the system with $\vectr{M}$ in the $y$-$z$ plane, and with the driving field along the $x$-axis. Define
\begin{equation}
\label{e.5}
\xi \equiv \left(\frac{\gamma h_d}{\omega}\right)\sin{\omega t}~.
\end{equation}
Then,
\begin{equation}
\label{e.6}
M_y(t) = M_{y0} \cos \xi + M_{z0} \sin \xi
\end{equation}
\begin{equation}
\label{e.7}
M_z(t) = M_{z0}\cos \xi - M_{y0} \sin \xi~.
\end{equation}
where $M_{y0}$ and $M_{z0}$ are the initial values for the components of the magnetization. Note that the motion in this limit, although nonlinear, is not chaotic.
\\
\subsection{Time evolution}
We now examine the time evolution of the system for different driving field amplitudes. The system will be driven at a frequency of $\omega = 1$ GHz and an initial position given by $M_{y0}=M_{z0}=M_s/\sqrt 2$ . Because the system is started in a nonequilibrium configuration, there will be a transient response before the system settles into equilibrium precession at a frequency which matches the driving field frequency The transient involves time-decay of precession at the natural frequency $\omega_0 = 0.29$ GHz, using eq.~(\ref{e.4}).

We have used a variety of numerical integration schemes to solve the differential equation for the motion of the magnetization, eq.~(\ref{e.1}). These include 2nd and 4th order Runge-Kutta and LSODA. All the techniques yielded equivalent results. Unless otherwise stated, the parameters for our calculation are $H_0=100$ Oe, $M_s = 1.67$ $\times10^3$ G (saturation magnetization appropriate for Fe), $\gamma = 1.83\times10^{10}$ rad/s, and $\alpha = 0.01$.

\begin{figure}
\centerline{\includegraphics[width=0.48\textwidth]{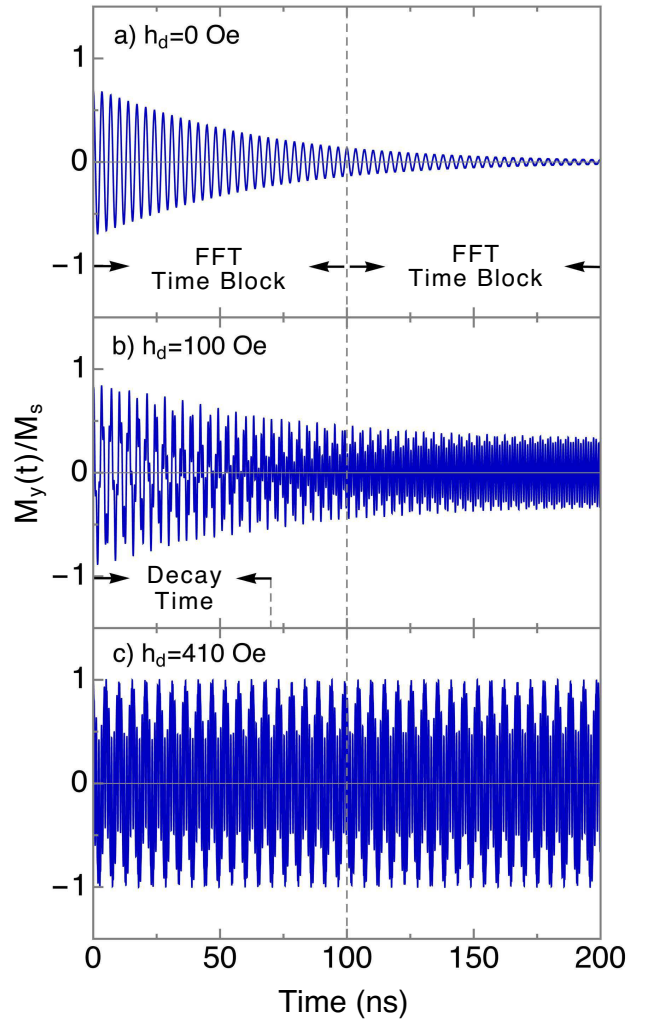}}
\caption{Time evolution of the normalized $M_y$ value for three different driving field amplitudes.  The natural frequency of the transient is 0.29 GHz and the frequency of the driving field is always 1 GHz. Fourier transforms have been applied in 100 ns intervals. For (a) and (b), the decay time of the transient is approximately 70 ns. For (c), the transient has shifted} slightly in frequency and has a much longer lifetime as the amplitude of the driving field is increased.
\label{f.1}
\end{figure}

In fig.~\ref{f.1}(a) where there is no driving field, we see a typical time evolution of damped oscillatory behavior. The decay time is about 70 ns. Figure~\ref{f.1}(b) shows the behavior of $M_y(t)$ with a moderate driving field of 100 Oe. In this limit, it should first be noted that the transient (natural) frequency is reduced very slightly from $\omega_0$. Second, one can also see that the system comes into equilibrium with the driving field frequency. Since the driving frequency is substantially higher than $\omega_0$, the system's evolution toward equilibrium is exhibited by the transition from the widely spaced time-trace (natural frequency behavior) toward the dense, closely packed region in the time-trace (driven high frequency behavior. The time it takes for the natural frequency to decay is once again about 70 ns.\\
Finally in fig.~\ref{f.1}(c), where a very strong driving field is applied, we see an interesting behavior where both the frequency of the transient (near $\omega_0$) and of the driving field ($\omega$) are seen.
\\
\begin{figure}
\centerline{\includegraphics[width=0.5\textwidth]{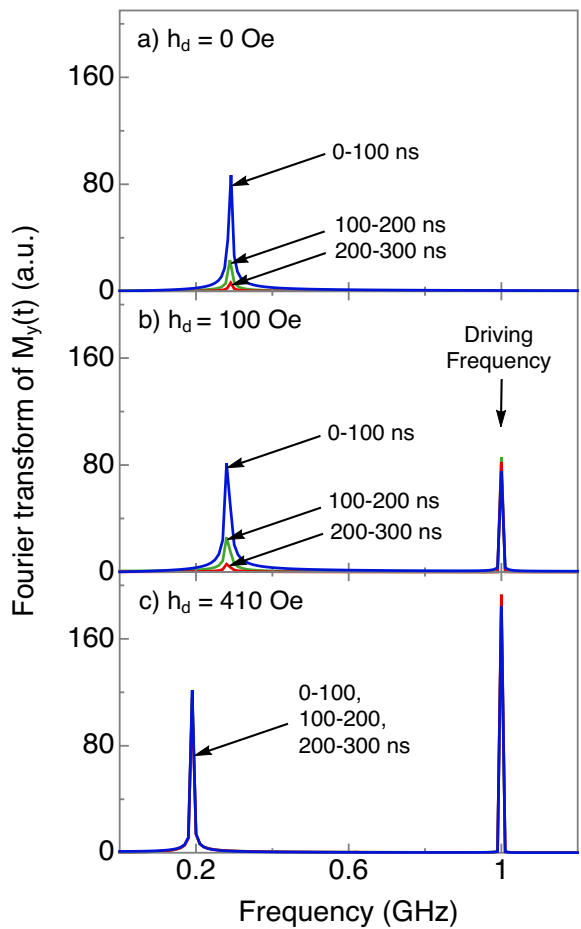}}
\caption{Fourier transform of $M_y(t)$ for different driving field amplitudes during different approximate time intervals.  In panels (a) and (b) we see the low frequency peak decays with time.  In panel (c) where the driving amplitude is large, the transient peak is essentially stabilized and does not decay in the time shown here.}
\label{f.2}
\end{figure}

\subsection{Fourier Analysis}
To quantify the transient time behavior of the magnetization, we do a series of Fourier transforms on $M_y(t)$, each transform encompassing a different time block (approximately 0--100 ns, 100--200 ns, etc).  These transforms will identify the major frequencies present in a given time block and the amplitude of the excitations at the different frequencies.  

The results are shown in fig.~\ref{f.2}.  In fig.~\ref{f.2}(a), we see the Fourier transform as a function of frequency when $h_d = 0$ Oe, for different time blocks. There is a single peak at the natural frequency, which diminishes in height as time is increased.  Figure~\ref{f.2}(b) shows the results when $h_d=100$ Oe. Now there are two peaks, one at the driving frequency and one very slightly below the natural frequency. Again the peak near the natural frequency is reduced as time is increased, indicating the typical decay in a damped linear system. The situation is quite different in fig.~\ref{f.2}(c) where the driving field is given by $h_d=410$ Oe. First, the frequency of the transient is shifted down further by the nonlinear interaction. More importantly, the height of the peak representing the transient now remains virtually constant over the time span investigated, indicating a substantially increased lifetime.
\\
\begin{figure}
\centerline{\includegraphics[width=0.5\textwidth]{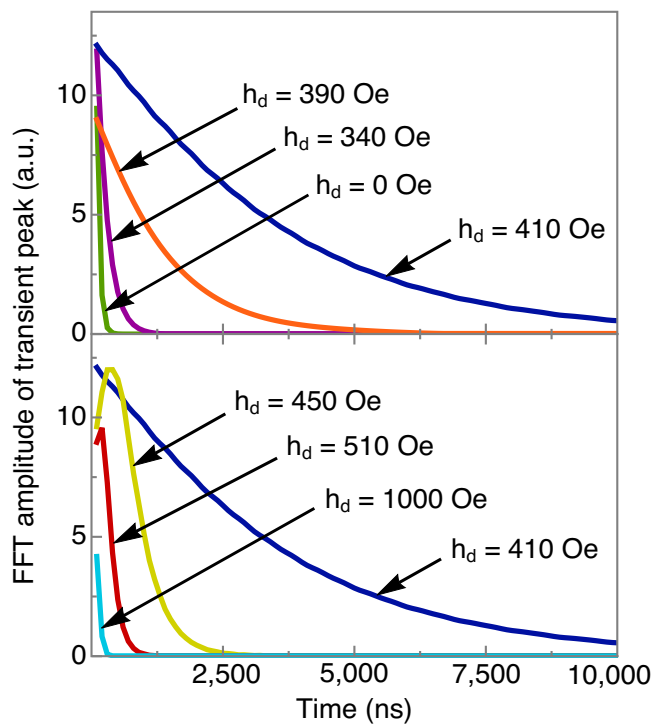}}
\caption{The decay time as a function of the amplitude of the driving field ($h_d$). In the upper panel the decay time \textbf{increases} as the driving field amplitude is increased to 410 Oe. In the lower panel the decay time \textbf{decreases} as the driving field amplitude is increased above 410 Oe.}
\label{f.3}
\end{figure}
\subsection{Amplitude decay of Fourier Transform}
The decay in time for different driving field amplitudes is illustrated in fig.~\ref{f.3}, where we plot the amplitude of the natural frequency mode as a function of time for different values of the driving field. Figure~\ref{f.3}(a) shows results for driving fields up to 410 Oe.  We see a remarkable and tunable increase in the duration of the transient, by about a factor of 100 for the range of values shown. When the driving fields are increased above the value of 410 Oe, fig.~\ref{f.3}(b), the transient lifetime is now reduced, returning to values close to that found in the linear case when $h_d$ is small.
\\ \indent The features illustrated here are quite robust.  For example, the general behavior of transient lifetime is essentially unaffected by all the following:
\\1) The initial position of the magnetization;
\\2) The initial phase of the driving field, i.e. the value of $\phi$ if the driving field is proportional to $\cos{\omega t+\phi}$;
\\3) The value of the damping constant, which scan even be zero.
\\4) Small changes in the structure, which cause a deviation from the spherical or cubic symmetry in the demagnetization tensor. (A sphere or cube has no nonlinear terms  involving the demagnetization tensor.  These terms do appear, however, for small changes in the structure, but they are not critical to the effect seen here.) 

In fact, the lifetime of the transient can be extended substantially by small variations in the amplitude of the driving field. In our example above, a driving field of 410 Oe gives a decay time of 1,300 ns.  Driving fields of 414 Oe and 414.7 Oe produce decay times of 23,600 and 1,480,000 ns respectively. We see a dramatic increase in lifetime as the driving field approaches the critical value.  
\\
\subsection{A change in equilibrium direction}
\begin{figure}
\centerline{\includegraphics[width=0.5\textwidth]{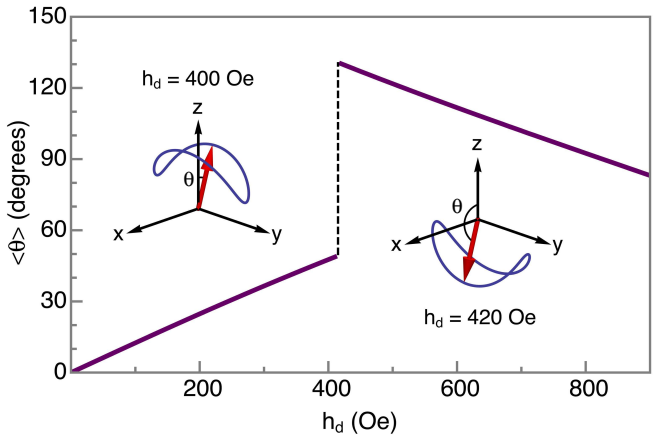}}
\caption{Plot of the average value of the polar angle, $\theta$, as a function of the driving field amplitude. The average is found after all transients have died down. Near the critical value of $h_d = 410$ Oe there is a transition where the   component of the magnetization changes from pointing along the $+z$ axis to pointing along the $-z$ axis.}
\label{f.4}
\end{figure}

\begin{figure}
\centerline{\includegraphics[width=0.5\textwidth]{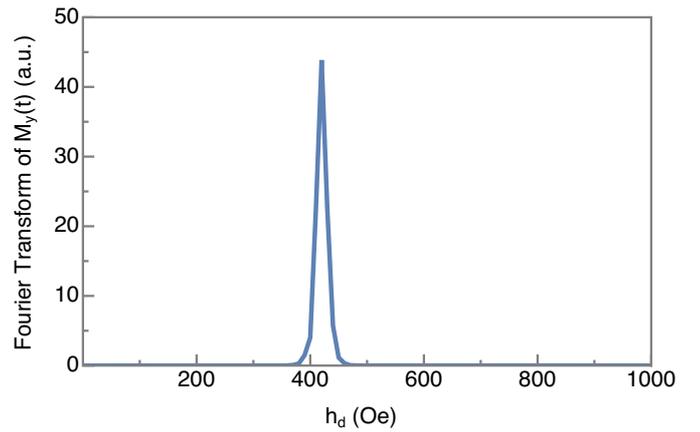}}
\caption{Fourier Transform of $M_y(t)$ as a function of the driving field, $h_d$. The transform is applied to a fixed interval of 2000--2100 ns. The peak amplitude of the Fourier Transform is centered around the critical driving field, $h_d \approx 410$ Oe, which is associated with a change in equilibrium direction.}
\label{f.5}
\end{figure}

We now explore the reason for the extended lifetime.  It is associated with a transition in the equilibrium direction induced by the driving field as seen in fig.~\ref{f.4}.  For low values of the driving field, the magnetization precesses around the static field, the $+\hat{\vectr z}$ direction. As the driving field is increased, the equilibrium direction eventually shifts to values having a component along the $-\hat{\vectr{z}}$ direction. The dramatic effect that the change in orientation has on the amplitude of the transient can be seen in fig.~\ref{f.5}. Here we plot the FFT amplitude of $M_y(t)$ for the time interval of 2000--2100 ns. Away from the transition the amplitude is essentially zero. But there is a significant effect over a range of driving field amplitudes, from about $h_d = 380$ Oe to $h_d = 450$ Oe. 

The situation is similar in many respects to a driven pendulum. In that case, there is only one energy minimum, when the pendulum is at the bottom position. However, with a large driving field, the pendulum motion is nonlinear and it can oscillate about a new equilibrium position where the pendulum is at the uppermost position \cite{b.23,b.24}. The physical mechanism for the stabilization of the ``inverted pendulum" has had an extensive discussion in the literature \cite{b.25}. 
\\
\begin{figure}
\centerline{\includegraphics[width=0.5\textwidth]{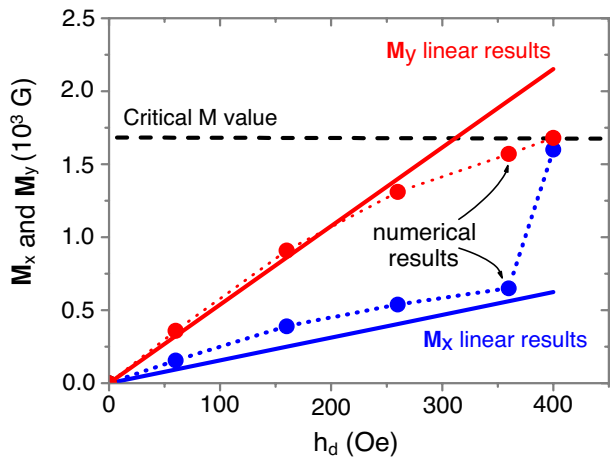}}
\caption{The transverse components of $\vectr M$ as a function of the driving field.  The solid lines indicate the predictions of linear theory, and the large dots show the results of a full nonlinear numerical calculation. The condition $\vectr M_y \approx \vectr M_s$ gives a reasonable approximation for the driving field at which the equilibrium direction changes.}
\label{f.6}
\end{figure}
\subsection{Finding the critical field}
We can develop an approximate analytic expression for the critical field necessary for this phase transition. Because the transition involves a change from precession in the upper half space ($+\hat{\vectr z}$) to the lower half space ($-\hat{\vectr z}$) one might expect that the transition occurs near a critical amplitude, the point at which the average value of $M_z$ becomes zero, or equivalently where the magnitude of $M_y$ is equal to $M_x$. We can estimate this using the linearized equations of motion for $M_x$ and $M_y$. One finds by solving eqs.~(\ref{e.1}) and (\ref{e.2})  
\begin{equation}
\label{e.8}
M_x = \left(\frac{\gamma^2 H_0 M_s}{\gamma^2 H_0^2-\omega^2}\right)h_d
\end{equation}
\begin{equation}
\label{e.9}
M_y = i\left(\frac{\gamma M_s \omega}{\gamma^2 H_0^2 - \omega^2}\right)h_d
\end{equation}
We neglect the $\gamma H_0$ in the denominator because it is small compared to the driving frequency $\omega$ here. Setting the magnitude of $M_y$ equal to $M_s$ we obtain the condition for the critical driving field
\begin{equation}
\label{e.10}
h_c = \frac{\omega}{\gamma} .
\end{equation}
With our parameters, we find the expected transition field to be about 340 Oe, somewhat lower than the 410 Oe we find numerically. 

It is somewhat surprising that this estimate works as well as it does. One reason for this is that the numerical calculations (shown in fig.~\ref{f.5} by dots) demonstrate that eqs. (\ref{e.8}) and (\ref{e.9}) are reasonable predictors of the maximum amplitudes of $M_x$ and $M_y$ (shown in fig.~\ref{f.5} by lines) for values of $h_d$ which are as large as 275 Oe.  The analytical form can explain some general trends we find in the numerical data. For example, we find that doubling the driving frequency requires a doubled critical field amplitude to reach the longest lifetimes, consistent with eq. (\ref{e.10}). We note that $M_x$ is smaller than $M_y$ because the driving field is in the $\hat{\vectr x}$ direction, so the critical condition that the linear prediction for $M_y\approx M_s$ is the correct one to use, as it will occur at a lower driving field. If the driving frequency were smaller than the resonance frequency the value of $M_x$ would generally be larger than $M_y$ as can be seen from eqs. (\ref{e.8}) and (\ref{e.9}).

The idea that the transient lifetime is tunable due to a change in equilibrium direction explains why this lifetime is insensitive to so many parameters. We drive the system off-resonance, so damping is not important. Furthermore, the transition occurs near a large amplitude precession, but the amplitude (in the linear approximation) is independent of initial phase and initial position. We note that the rapid transition from one equilibrium state to another seems to occur only for the case where the driving frequency is larger than the natural frequency.
\\
\section{Conclusion}
The general idea developed here is that a transient oscillation at a natural frequency may be stabilized by a driving field at a different frequency, particularly near a change in equilibrium position. This phenomenon could be detected using the micro-SQUID experiment of ref.~\cite{b.22}. The concept of extending the lifetime of a transient can occur in the chaotic regime \cite{b.26,b.27,b.28}, but here we have developed the result in a \textit{nonchaotic} region. The slow decay of the transient is surprising.  In part it seems to be associated with fact that the entire structure takes a long time to settle into equilibrium position near the sharp transition.  However, our numerical calculations show that the transient decay time is much longer than the equilibration time.

It is interesting to note that similar results are not found in other nonlinear physical systems such as the nonlinear driven pendulum in one dimension, the spherical pendulum, or nonlinear spring systems (with nonlinear forces that are either quadratic of cubic in the displacement). Although these cases do exhibit a transition from one equilibrium state to another, the lifetime of the natural frequency mode does in fact not change. Thus our results naturally lead us to the intriguing question - what makes the nonlinear magnetization equations unique relative to many other nonlinear systems? 

\acknowledgments
This work was supported by UCCS Letters Arts and Sciences Faculty/Student Research Grant.


\begin{thebibliography}{0}

\bibitem{b.1}
  \Name{Wiggen P. E.}
  \Book{Nonlinear Phenomena and Chaos in Magnetic Materials}
  \Publ{World Scientific, Singapore}
  \Year{1994}.
  
\bibitem{b.2}
  \Name{Bertotti G., Mayergoyz I. D. \and Serpico C.}
  \Book{Nonlinear Magnetization Dynamics in Nanosystems}
  \Publ{Elsevier Ltd, Amsterdam}
  \Year{2009}.

\bibitem{b.3}
  \Name{Wu M.}
  \Book{Solid State Phys.}
  \Editor{Camley R. E. \and Stamps R. L.}
  \Vol{62}
  \Publ{Academic Press, Burlington}
  \Year{2011}
  \Pages{163}{224}.
  
\bibitem{b.4}
  \Name{Kosevich A. M., Ivanov B. A. \and Kovalev A. S.}
  \REVIEW{Phys. Rep.}{194}{1990}{117}.

\bibitem{b.5}
  \Name{Slavin A. N. \and Rojdestvenskii I. V.}
  \REVIEW{IEEE Trans. Magn.}{39}{1994}{37}.
  
\bibitem{b.6}
  \Name{Wu M., Kalinikos B. A. \and Patton C. E.}
  \REVIEW{Phys. Rev. Lett.}{95}{2005}{237202}.

\bibitem{b.7}
  \Name{Demokritov S., Serga A. A., Demidov V. E., Hillebrands B., Kostylev M. P. \and Kalinikos A.}
  \REVIEW{Nature}{426}{2003}{159}.
  
\bibitem{b.8}
  \Name{Smith R. K., Grabowski M. \and Camley R. E.}
  \REVIEW{J. Magn. Magn. Mater.}{332}{2010}{2127}.
  
\bibitem{b.9}
  \Name{Marsh J. \and Camley R. E.}
  \REVIEW{Phys. Rev. B.}{86}{2012}{224405}.
  
\bibitem{b.10}
  \Name{Cheng C. \and Bailey W. E.}
  \REVIEW{Appl. Phys. Lett.}{103}{2013}{242402}.
  
\bibitem{b.11}
  \Name{L'vov V. S.}
  \Book{Wave Turbulence Under Parametric Excitation}
  \Publ{Springer-Verlag, Berlin}
  \Year{1994}.
  
\bibitem{b.12}
  \Name{Berkov D. \and Gorn N.}
  \REVIEW{Phys. Rev. B.}{71}{2005}{052403}.
  
\bibitem{b.13}
  \Name{Murugesh S. \and Lakshmanan M.}
  \REVIEW{Chaos}{19}{2009}{043111}.
  
\bibitem{b.14}
  \Name{Khivintsev Y., Marsh J., Zagorodnii V., Harward I., Lovejoy J., Kirvosik P., Camley R. E. \and Celinksi Z.}
  \REVIEW{Appl. Phys. Lett.}{98}{2011}{042505}.
  
\bibitem{b.15}
  \Name{Wismayer M. P., Southern B. W., Fan X. L., Gui. Y. S., Hu C. -M. \and Camley R. E.}
  \REVIEW{Phys. Rev. B.}{85}{2012}{064411}.

\bibitem{b.16}
  \Name{Kostylev M., Demidov V. E., Hanson U. -H. \and Demokritov S. O.}
  \REVIEW{Phys. Rev. B.}{76}{2007}{224414}.
  
\bibitem{b.17}
  \Name{Ulrichs H., Demidov V. E., Demokritov S. O. \and Urazhdin S.}
  \REVIEW{Phys. Rev. B.}{84}{2011}{094401}.  
  
\bibitem{b.18}
  \Name{Nembach H. T., Livesey K. L., Kostylev M. P., Martin-Pimentel P., Hermsdoerfer S. J., Leven B., Fassbender J. \and Hillebrands B.}
  \REVIEW{Phys. Rev. B.}{84}{2011}{184413}.
  
\bibitem{b.19}
  \Name{Yang Z., Zhang S. \and Li Y. C.}
  \REVIEW{Phys. Rev. Lett.}{99}{2007}{134101}.
  
\bibitem{b.20}
  \Name{Laroze D., Becerra-Alonso D., Gallas J. A. C. \and Pleiner H.}
  \REVIEW{IEEE Trans. Magn.}{48}{2012}{3567-3570}.
 
 \bibitem{b.21}
  \Name{Coey J. M. D.}
  \Book{Magnetism and Magnetic Materials}
  \Publ{Cambridge University Press, Cambridge}
  \Year{2009}
  \Page{361}.
  
\bibitem{b.22}
  \Name{Thiron C., Wernsdorfer W. \and Mailly D.}
  \REVIEW{Nat. Mater.}{2}{2003}{524}.
  
\bibitem{b.23}
  \Name{Kapitza P. L.}
  \REVIEW{Sov. Phys. J. Exp. Theor. Phys.}{21}{1951}{588}.
  
\bibitem{b.24}
  \Name{Housner G. W.}
  \REVIEW{Bull. Seismol. Soc. Am.}{53}{1963}{403}.
  
  \bibitem{b.25}
  \Name{Butikov E. I.}
  \REVIEW{Am. J. Phys}{69}{2001}{1}.
  
 \bibitem{b.26}
  \Name{Lai Y. C.}
  \Book{Nonlinear Dynamics and Chaos: Advances and Perspectives}
  \Editor{Thiel M., de Moura A. P. S. \and Grebogi C.}
  \Publ{Springer-Verlag, Berlin}
  \Year{2010}
  \Pages{131}{152}.
  
\bibitem{b.27}
  \Name{Lin Y. -Y., Lisitza N., Ahn S. \and Warren W. S.}
  \REVIEW{Science}{290}{2000}{181}.
  
\bibitem{b.28}
  \Name{Datta S., Huang S. Y. \and Lin Y. -Y.}
  \REVIEW{J. Chem. Phys.}{124}{2006}{154501}.

\end{thebibliography}
\end{document}